
%
%
\input harvmac
\def\Title#1#2{\rightline{#1}\ifx\answ\bigans\nopagenumbers\pageno0\vskip1in
\else\pageno1\vskip.8in\fi \centerline{\titlefont #2}\vskip .5in}

%
%

\def\caln{{\cal N}}
\def\lpl{l_{\rm pl}}
\def\Mpl{M_{\rm pl}}

\def\eg{{\it e.g.}}

\def\ket#1{\vert#1\rangle}
\def\bra#1{\langle#1\vert}

\font\ticp=cmcsc10

\def\ajou#1&#2(#3){\ \sl#1\bf#2\rm(19#3)}

\def\Hsl{{\,\raise.15ex\hbox{/}\mkern-12mu H}}
\def\Dsl{\,\raise.15ex\hbox{/}\mkern-13.5mu D}
\def\Ssl{{\,\raise.15ex\hbox{/}\mkern-10.5mu S}}

\noblackbox
%
%
\lref\HawkETD{S.W. Hawking, ``Evaporation of two dimensional black holes,''
CalTech preprint CALT-68-1774, hepth@xxx/9203052}
\lref\QBH{B. Birnir, S.B. Giddings, J.A. Harvey, and A. Strominger,
``Quantum black holes,'' UCSB/Chicago preprint UCSB-TH-92-08=EFI-92-16,
hepth@xxx/\-9203042.}
\lref\SuTh{L. Susskind and L. Thorlacious, ``Hawking radiation and
back-reaction,'' Stanford preprint SU-ITP-92-12, hepth@xxx/9203054.}
\lref\NSSW{V.P. Nair, A. Shapere, A. Strominger, and F. Wilczek,
``Compactification of the twisted heterotic string,''\ajou Nucl. Phys.
&B287 (87) 402.}
\lref\BPS{T. Banks, M.E. Peskin, and L. Susskind, ``Difficulties for the
evolution of pure states into mixed states,''\ajou Nucl. Phys. &B244 (84)
125.}
\lref\HawkUnc{S.W. Hawking, ``The unpredictability of quantum
gravity,''\ajou Comm. Math. Phys &87 (82) 395.}
\lref\Dual{M.B. Green, J.H. Schwarz, and L. Brink, ``N=4 Yang-Mills and
N=8 supergravity as limits of string theories''\ajou Nucl. Phys. &198B (82)
474\semi
K. Kikkawa and M. Yamasaki, ``Casimir effects in superstring
theories,''\ajou Phys. Lett. & 149B (84) 357\semi
N. Sakai and I. Senda, ``Vacuum energies of string compactified on
torus,''\ajou  Prog. Theor. Phys. Suppl. &75 (86) 692.}
\lref\Haij{P. H\'aji\v cek, ``Quantum mechanics of gravitational
collapse,'' Berne preprint BUTP-92/4.}
\lref\Wald{R.M. Wald, ``Black holes, singularities and predictability,'' in
{\sl Quantum theory of gravity. Essays in honor of the 60th birthday of
Bryce S. Dewitt}, S.M. Christensen (Ed.),  Hilger (1984);
``Black holes and thermodynamics,'' U. Chicago preprint, lectures at 1991
Erice school on Black Hole Physics.}
\lref\Vene{G. Veneziano, ``An enlarged uncertainty principle from gedanken
string collisions?'' in {\sl Strings `89}, R. Arnowitt et. al. (Eds.), World
Scientific (1990).}
\lref\ACV{D. Amati, M. Ciafaloni, and G. Veneziano, ``Superstring
collisions at planckian energies,''\ajou Phys. Lett. &197B (87) 81;
``Classical and quantum gravity effects from planckian energy superstring
collisions,''\ajou Int. J. Mod. Phys. &A3 (88) 1615.}
\lref\Gros{D.J. Gross, ``Superstrings and Unification," in
{\sl Munich high energy physics}, R. Kotthaus and J.H. Kuehn (Eds.),
Springer (1989).}
\lref\tHoo{G. 't Hooft, ``The black hole interpretation of string
theory,''\ajou Nucl. Phys. &B335 (90) 138.}
\lref\DJT{N. Deo, S. Jain, and C.-I Tan, ``Strings at high-energy densities
and complex temperature,''\ajou Phys. Lett. &B220 (89) 125; ``String
statistical mechanics above the hagedorn energy density,''\ajou Phys. Rev.
&D40 (89) 2626.}
\lref\MiTu{D. Mitchell and N. Turok, ``Statistical mechanics of cosmic
strings,''\ajou Phys. Rev. Lett. &58 (87) 1577; ``Statistical properties of
cosmic strings,''\ajou Nucl. Phys. &B294 (87) 1138.}
\lref\GrMe{D.J. Gross and P. Mende, ``The high-energy behavior of string
scattering,''\ajou Phys. Lett. &197B (87) 129; ``String theory beyond the
Planck scale,''\ajou Nucl. Phys. &B303 (88) 407.}
\lref\LNW{K. Lee, V.P. Nair, and E.J. Weinberg, ``Black holes and magnetic
mono\-poles,'' Fermilab/Columbia preprint CU-TP-539=FERMILAB-Pub-91/312-A\&T;
``A classical instability of Reissner-Nordstr\"om solutions and the fate of
magnetically charged black holes,'' Fermilab/Columbia preprint
CU-TP-540=FERMILAB-Pub-91/326-A\&T.}
\lref\Polc{J. Polchinski, ``Decoupling versus excluded volume, or return of
the giant wormholes,''\ajou Nucl. Phys. &B325 (89) 619.}
\lref\ColeComm{S. Coleman, private communication.}
\lref\BoGi{M.J. Bowick and S.B. Giddings, ``High-temperature
strings,''\ajou Nucl. Phys. &B325 (89) 631.}
\lref\BDDO{T. Banks, A. Dabholkar, M.R. Douglas, and M O'Loughlin, ``Are
horned particles the climax of Hawking evaporation?'' Rutgers preprint
RU-91-54.}
\lref\DXBH{S.B. Giddings and A. Strominger, ``Dynamics of Extremal Black
Holes,'' UCSB preprint UCSB-TH-92-01, hepth@xxx/9202004.}
\lref\PresComm{J. Preskill, private communication.}
\lref\GiStInc{S.B. Giddings and A. Strominger, ``Loss of incoherence and
determination of coupling constants in quantum gravity,''\ajou Nucl. Phys.
&B307 (88) 854.}
\lref\ColeHerr{S. Coleman, ``Black holes as red herrings: Topological
fluctuations and the loss of quantum coherence,''\ajou Nucl. Phys. &B307
(88) 867.}
\lref\HawkEvap{S.W. Hawking, ``Particle creation by black
holes,"\ajou Comm. Math. Phys. &43 (75) 199.}
\lref\Dyso{F. Dyson, Institute for Advanced Study preprint, 1976,
unpublished.}
\lref\GPS{S.B. Giddings, J. Preskill, and A. Strominger, unpublished.}
\lref\ACN{Y. Aharonov, A. Casher, and S. Nussinov, ``The unitarity
puzzle and Planck mass stable particles,"\ajou Phys. Lett. &B191 (87)
51.}
\lref\CGHS{C. Callan, S.B. Giddings, J.A. Harvey, and A. Strominger,
``Evanescent Black Holes,"\ajou Phys. Rev. &D45 (92) R1005.}
\lref\DXBH{S.B. Giddings and A. Strominger, ``Dynamics of extremal
black holes," UCSB preprint UCSB-TH-92-01, hepth@xxx/9202004, to appear in
{\sl Phys. Rev. D.}}
\lref\Frau{S. Frautschi,
``Statistical Bootstrap Model of Hadrons,"\ajou Phys. Rev. &D3
(71) 2821.}
\lref\Carl{R.D. Carlitz,
``Hadronic Matter at High Density,"\ajou Phys. Rev. &D5 (72) 3231.}
\lref\Morg{D. Morgan, ``Black holes in cutoff gravity,"\ajou Phys.
Rev. &D43 (91) 3144.}
\Title{\vbox{\baselineskip12pt\hbox{UCSBTH-92-09}\hbox{hepth@xxx/9203059
}
}}
{\vbox{\centerline{Black Holes and Massive Remnants}
}}

\centerline{{\ticp Steven B. Giddings
}}
\vskip.1in
\centerline{ Department of Physics}
\centerline{ University of California}
\centerline{ Santa Barbara, CA 93106}
\bigskip
\centerline{\bf Abstract}
This paper revisits the conundrum faced when one attempts to
understand the  dynamics of black hole formation and
evaporation without abandoning unitary evolution.
Previous efforts to resolve this puzzle assume
that information escapes in corrections to the Hawking process, that an
arbitrarily large amount of information is transmitted by a
planckian energy or contained in a Planck-sized remnant, or that the
information is lost to another universe.  Each of these possibilities has
serious difficulties.
This paper
considers another alternative: remnants that carry
large amounts of information and whose size and mass depend on their
information content.  The existence of such objects
is suggested by attempts to incorporate
a Planck scale cutoff into physics.  They would greatly alter the
late stages of the evaporation process.  The main drawback of this scenario
is apparent acausal behavior behind the horizon.

\Date{3/92}
\newsec{The Conundrum}
Consider a pure quantum state
corresponding to a distribution of matter of
mass $M$ collapsing due to its gravitational self attraction.  Such a
state can be described by a density matrix $\rho=\ket\psi\bra\psi$
with vanishing entropy, $S=-{\rm Tr}\rho\ln\rho$.  For sufficiently
large $M$ it is believed that the generic such state will eventually
pass its event horizon and form a black hole.  According to
Hawking\refs{\HawkEvap},
the resulting black hole will then lose mass by evaporation.  It is
believed that this process can be treated using Hawking's semiclassical
calculation up until quantum-gravitational and back-reaction
effects become important.  One expects this to happen when
the mass of the black hole becomes comparable to the Planck mass, $\Mpl$.

Once the black hole has evaporated to $M\sim\Mpl$, most of
the initial mass is contained in outgoing Hawking radiation.  At
any instant in the evaporation process this radiation is approximately
described
by a
thermal density matrix with a non-zero entropy.  The total entropy in
the outgoing state is estimated to be
$S\sim M^2/\Mpl^2$.  Since entropy is absence of information,
its nonvanishing indicates that information
that was contained in
the initial quantum state and has fallen into the black hole does
not subsequently escape in the Hawking process.  The information
problem for black holes is the problem of explaining what happens to
this missing information.

Since Hawking's initial results several possible fates have been proposed
for information
lost to a black hole.\foot{For another nice discussion of some aspects of
the problem see \refs{\Wald}.}
The first possibility is

\item{A.}  {\it The black hole evaporates completely, and the information
contained within it is irretrievably lost.}

This option implies that the
evolution of the complete system, including the black hole, is
fundamentally nonunitary\refs{\HawkUnc}.  Although there is no known basic
inconsistency in this scenario, and hence it is possibly right, it
is clearly disturbing. Furthermore,
there are general arguments\refs{\BPS} that such non-unitary
evolution implies violation of either locality or energy-momentum
conservation.

For these reasons, in the rest of the paper it will
be assumed that this possibility is not correct and alternatives will
be sought.

A second option is originally due to Dyson\refs{\Dyso}:

\item{B.} {\it The black hole disappears completely, but one or more
separate universes branch off during the process and carry away the
information. }

This proposal appears to neatly accommodate the apparent nonunitarity of the
Hawking process while preserving fundamental unitarity.  It does so by
enlarging
the Hilbert space to include states on the separate universes.  However, it
has been difficult to find a concrete model producing such a result.
Furthermore, there is an apparent issue of principle:  it seems that arguments
similar to those used for baby universes\refs{\ColeHerr,\GiStInc} can
be used to replace the effects of the other universes by
arbitrariness in coupling constants in an otherwise unitary theory
describing evolution in
our universe.  If so,
one is still left with the need to explain
information lost to the black hole in the resulting unitary
theory\refs{\ColeComm,\GPS}.

Two other possibilities raise the question of whether corrections to
the Hawking calculation reinstate the missing information.  The first
of these is:

\item{C.}  {\it The black hole disappears completely, and the information
is transmitted from the infalling matter to the outgoing Hawking
radiation by higher order effects neglected in the original
calculation.  }

This scenario, which has been advocated for example by 't Hooft\refs{\tHoo},
would require that the Hawking radiation
extracts {\it all} information from the ingoing matter, \eg\ through scattering
as they cross near the horizon.  In particular, matter that crosses the
horizon and falls towards $r=0$ must therefore have essentially zero
information
content.  For this reason this option seems
seems farfetched to many.

A slightly different alternative  is:

\item{D.}  {\it As the radius of the horizon gradually shrinks,
information in the internal state that was previously hidden by the
horizon is gradually revealed.  When one reaches the Planck radius,
essentially all of the information has been released in this
fashion.}\foot{This
possibility was pointed out to me by K. Kucha\v r.}

Due to the tendency of particles, hence information, to focus towards
$r=0$, it is questionable whether sufficient information can be shown
to escape in this way.  Furthermore, one could imagine a process where a
continuous flux of incoming coherent radiation precisely balances the outgoing
Hawking flux, in which case the horizon doesn't shrink and an unbounded
amount of information is lost.\foot{I learned this argument from Y.
Aharonov.}

The final two options involve the details of the endpoint of evaporation:

\item{E.}  {\it In the late stages of evaporation, during which the energy
equivalent of the last
few Planck masses is emitted, and where the Hawking calculation breaks down,
the new dynamics that replaces it
allows all of the missing information to be
emitted. }

This last option requires that an unboundedly large amount of
information (of order $\exp[{M^2/\Mpl^2}]$, where $M$
is the mass of the
initial state, which can be arbitrarily large) be
carried by an energy the equivalent of several Planck
masses.  Locality, causality and energy conservation put rather
stringent constraints on how this may happen.

To see this\refs{\ACN}, note that in order to make up the entropy
$S\sim M^2/\Mpl^2$, of order $\caln = M^2/\Mpl^2$ light quanta must
be emitted during the last few Planck masses of energy emission.  The
average wavelength of each such quantum is then of order $\lambda\sim
\caln\lpl$ where $\lpl$ is the Planck length.  This is to be compared
with the final size of the black hole, $R_{\rm bh}\sim \lpl$.  The
very small wavefunction overlap then contributes a suppression factor
$1/\caln^3$ to the emission probability.  This means that if the final
decay must occur by simultaneous emission of all $\caln$ quanta, then
the lifetime for the final decay is comparable to the age of the
Universe  for initial masses a few times the Planck
mass.  Even if one assumes that the final decay takes place by
gradual emission of the $\caln$ quanta, the lifetime is of
order\foot{Preskill has obtained a different bound by thermodynamic
arguments\refs{\PresComm}. }
$\tau \sim \caln^4 t_{\rm pl}$, where $t_{\rm pl}$ is the Planck
time.  This lifetime is of order the age of the Universe for
even kilogram-sized black holes.
Therefore one is forced to conclude that there are long-lived
remnants with Planck-sized  masses that carry unbounded amounts of
information.  One can then restate this as a sixth proposal:

\item{F.}  {\it The end-product of the Hawking evaporation is a long-lived
remnant with mass comparable to $\Mpl$ and which carries arbitrarily
large amounts of information corresponding to an infinite number
of internal states.}

There are also potential problems with this latter possibility.
First, this infinite variety of Planck mass particles will appear in loops
and in the thermodynamic ensemble.  In either case, there appear to be serious
difficulties
caused by the infinite degeneracy.
Furthermore, it is questionable whether it is physically
reasonable for an unbounded amount of information to be carried
within a Planck volume.\foot{Note however, that one explanation for
how this could appear to
happen was proposed in \refs{\CGHS\BDDO-\DXBH}
within the
context of charged dilatonic black holes.  For these it was
argued that inside what seems to an outside observer to be a small volume
surrounding the black hole
there is in fact an infinite-volume tube capable of storing infinite
information.  This point will be discussed
further in section 3.}

Although one of the alternatives A-F could conceivably be the key to
the black hole information problem, each appears to have sufficient
internal difficulties and/or conflicts with basic principles to render it
likely nonviable.  In the following we shall explore yet another
alternative that can escape some of these difficulties.
%
%
%

\newsec{Cautionary Note}

The remainder of this paper rests on speculations about
what may be the plausible behavior of gravity at short
distances.  Clues about this come from string theory, but even within the
context of string theory these speculations cannot be confirmed with our
present state of knowledge.

Nonetheless, it is felt that it is constructive to apply such
speculations to black hole theory and see what the logical outcome may be.
The reason for this is two-fold.

First,
we have seen that it is quite difficult to understand the process of
formation and evaporation of a black hole without either abandoning
cherished principles such as unitarity and locality or without
making various very
questionable and potentially unphysical assumptions.
The difficulty in arriving at such a plausible picture implies that
we must at the present investigate even very
speculative scenarios such as B,E,F, or others resting on
reasonable assumptions that nonetheless cannot presently be verified.
Such investigation should allow us to sharpen our
understanding of the constraints on such a scenario.  It may also
provoke the development of even more precise and reasonable
solutions to the black hole information problem.
In short, because of the difficulty of imagining reasonable resolutions to
the information problem, it is
imperative that we leave no stone unturned in the search for
even a qualitative
picture of black hole evaporation that allows us to retain our principles.

Second, the exploration of possible resolutions to the black hole
information problem can act as a guide to the behavior of the
microphysics of gravity.  If the only imaginable solutions to the problem
require particular
assumptions about the properties of that microphysics, then this is
impetus to believe that the assumed properties should arise in a
realistic theory of gravity.
One should then attempt
to see how candidates for the  microphysics could consistently
produce the specified behavior.
For example, one might seek to extract the required dynamics
from a fundamental theory such as string theory.

\newsec{Information Bounds and Black Hole Cores}

We begin by returning to one of the objections to option F:
it seems physically improbable that
an unbounded amount of information could be
contained in a volume of Planck size.  One should then ask what would be
a reasonable bound for information or entropy content.

There is precedent for limits on information content.
Within the context of classical mechanics, there is no bound to the
number of configurations that may correspond to a given volume of phase
space.  Quantum mechanics changes this through the uncertainty
principle.  It does so
by introducing
a  natural unit,
$\hbar$,
for the phase space volume element.
The statement that in $d$ dimensions there  can be only
one quantum state per unit $\hbar^d$ volume  restricts the
allowed information content.

One can likewise ask if, within the context of gravity, there might
be any similar constraints on information content.
On dimensional grounds the natural presumption is that there is
only one state per Planck volume.

There are physical reasons to suspect the existence
of such a bound.  In particular, there
is the general fact that at distances shorter
than the Planck length, spacetime as we know it should cease to exist.
One therefore believes that it is meaningless to consider excitations
whose wavelength is shorter than the Planck length.  Whatever serves
as the physical Planck-scale cutoff for gravity should eliminate
these excitations.

More specific arguments can be made within the context of a specific
theory of quantum gravity such as string theory.  In string theory
the cutoff at the Planck length is provided by the string scale.
String theory presents
various pieces of indirect evidence  that
there are not states at shorter distances.

The first piece of evidence comes from studies of the free string
gas.  When the entropy density reaches the planckian density, (or
equivalently the energy density reaches the Hagedorn density),
attempts to force more energy into the system lead to delocalization
of the string ensemble.  Within the context of strings in a
compactified space, this is evidenced by the production of winding
modes\refs{\MiTu\DJT-\BoGi} which wrap around the compact
dimensions.  In non-compact space, similar attempts again lead to
delocalization since the energy is forced into long
string\refs{\Frau\Carl,\MiTu\DJT-\BoGi}.  One can
ascribe these phenomena to string theory's attempt to avoid
super-planckian energy
and entropy densities.

A second piece of evidence arises in studies of high-energy string
scattering \refs{\GrMe,\ACV}.  It has been shown that  one
cannot explore distances shorter than the Planck distance using strings
as probes. When one increases the incident string energy in an
attempt to do so, instead the string probe
delocalizes upon interaction
with the target.  This has lead to the suggestion of a new ``string
uncertainty principle\refs{\Gros,\Vene}," in addition to the
traditional uncertainty principle of quantum mechanics.  As in
quantum mechanics this indicates the absence of degrees of freedom in
the theory; here these are the degrees of freedom on scales below the
string scale, or
Planck length.  This implies that information cannot be stored in
states below that scale.

Other evidence for absence of short-distance degrees of freedom
arises from duality symmetry. 	In its simplest form\refs{\Dual,\NSSW}
this states that
a string moving on a circle of radius $r$ is equivalent to a string
moving on a circle of radius $1/r$.  This has been generalized to
much broader contexts, and many believe it to be part of a
fundamental symmetry that removes short-distance degrees of freedom
in string theory.

Let us therefore take as our basic hypothesis, motivated either by general
considerations of quantum gravity or by string theory, that there is
an
upper bound on the information content within a given volume, and that this
bound is
determined by the cutoff at the Planck scale.
This has an immediate consequence for the black hole problem:
the entropy corresponding to that in a black hole formed from an initial
mass $M$ requires a volume $V_M$ that grows with $M$ to contain it.
One obvious guess is that the allowable
entropy density is bounded by the Planck density, which gives the estimate
$V_M\sim \lpl^3 \left(M/\Mpl\right)^2$.

This indicates that if one considers the initial black hole of mass $M$,
it  should have a finite sized core
in which information is distributed instead of being concentrated at
the singularity.  One might expect that this has volume of order $V_M$,
and a  na\"\i ve estimate of the the radius of the core region is therefore
\eqn\ebound{r_M\sim \lpl (M/\Mpl)^{2/3}\ .  }
In fact, a different argument
involving the need to cut off gravity at the Planck scale implies a
similar
result, albeit with a core radius mass dependence different
from the above estimate.
In \refs{\Morg}, Morgan considers black holes in a
model for gravity due to Polchinski\refs{\Polc} in which it is
assumed that there is an upper bound on the eigenvalues of the
curvature tensor,
\eqn\curvbd{\vert R\vert \leq {1/\Lambda^2}}
where $\Lambda$ is a cutoff of order $\lpl$.
An example of a non-trivial curvature invariant in the Schwarzschild
solution is
\eqn\curv{R_{\alpha\beta\gamma\delta}R^{\alpha\beta\gamma\delta} =
48{M^2\over r^6}\ .}
Combining this with \curvbd\ implies that the region inside
radius
\eqn\cbound{r\sim \lpl \left(M/\Mpl\right)^{1/3}}
should be removed;  Morgan argues that it is
replaced by a singularity-free core.

Although the qualitative picture  of a large core, rather than a
singularity, seems robust, one might have hoped for the same mass dependence
from both the entropy estimate \ebound\ and the curvature estimate \cbound\
of the core
radius.  However, one
should note that it is not understood precisely how the cutoff should
appear in quantum gravity, and in particular a simple cutoff on
the entropy density or on the curvature is probably too na\"\i ve.
Furthermore, there are two different logical possibilities for how a core
with size of order $V_M$ can be accommodated within the black hole.  The first
is that the core extends out to some radius that grows with the mass, for
example as given by the entropy estimate above. A second possibility
is suggested by the work of refs.~\refs{\BDDO,\DXBH}:
in curved space an arbitrarily large volume can be hidden within a fixed
radius.  Thus the volume of the core may grow as its radius stays fixed;
the core
can be thought of as a large internal geometry
attached to the outside geometry through a
fixed size neck as in \fig\one{In curved space, an arbitrarily large amount
of information may be contained within a large volume that nonetheless
appears to external observers to have planckian size.}.
Gravity with a curvature cutoff presents a
picture that includes both of these possibilities.   The radius grows with the
mass, but the internal solution found in \refs{\Morg} is similar to an
expanding deSitter universe with cosmological constant given
by the cutoff scale.

Now consider the evolution of the core as the black hole evaporates.
In accord with the arguments of the preceding section we assume that the
information does not escape in the Hawking radiation or through topology
change.  If one takes the curvature bound for the core radius, then the
core radius shrinks with the mass of the evaporating black hole.
However, since the core itself
must act as a repository of information one expects
it not to shrink. Indeed, in the model of \refs{\Morg} it grows, albeit in
a highly cutoff-dependent way.  Once the radius of the horizon has shrunk
down to the Planck size, the horizon and neck meet.  It seems that there
are two possible results.
One is that the neck can pinch off; the core
then becomes a child universe.  This possibility returns us to scenario B
and its attendant difficulties.  Another possibility is that the core stays
connected to our universe through the neck.  In that case, one views it as
a Planck-sized remnant from the outside; the large interior of the core
contains the missing information but is not visible without passing
through the Planck-sized neck.  One is then back at option F.  While one
has eluded the prejudice against unbounded information densities, the
problems with loops and thermodynamics would still seem to
occur,\foot{I thank A.
Strominger and J. Preskill for discussions on this issue.} although it is
remotely conceivable that they could be cured by suppressed
amplitudes for creating such large cores if these
amplitudes decrease sufficiently
rapidly  with the
information content.  So in either case, we have been returned to the
scenarios discussed in the introduction and to their associated difficulties.

\newsec{Massive Remnants}

If one accepts the proposal that information density is bounded but remains
concerned about the problems with scenarios B and F, one is left with the
possibility that the radius of the core does not shrink to the Planck scale
as the black hole evaporates.  The persistence of such a finite size core
indicates both a revision of the Hawking scenario as well as a
resolution of the information problem.  Assume for example that the core
radius $r_M$
is fixed by the initial mass, in accord with the notion that the
core is of finite size to accommodate the bound on information
density.  In that case the Hawking evaporation should proceed
normally only until the point where the horizon size has shrunk to
$r_M$.  Then the core begins to protrude and produces a remnant.
Subsequently the standard
Hawking
calculation is no longer valid.\foot{A {\it superficially} similar
modification of the Hawking calculation has recently been noted in
the context of a magnetically charged black hole in the Higgs
model\refs{\LNW}.  There it was shown that once the hole evaporates
to a certain radius, a neighborhood of the horizon makes a transition
to the state corresponding to the monopole core of the
t'Hooft-Polyakov monopole.  The Hawking calculation for the
broken-symmetry vacuum is then no longer valid due to the presence of
the core.}

The radical difference between this scenario and traditional Hawking
process therefore
offers another resolution of the information problem:

\item{G.} {\it The Hawking process terminates with the production of a
large, massive remnant which carries all of the  information
missing from the outgoing radiation.  Whether the remnant subsequently
decays or whether it is long lived, the information stored in it becomes
accessible.}

Indeed,
since the
Hawking calculation is only valid until the core radius is reached,
the statement that the lost information is not reemitted is only
correct until this time.  After this time the core comes into causal
contact with the external world.  The subsequent evolution
of the remnant after the horizon has reached
the core radius would seem to depend on the details of the physics
introducing the planckian cutoff.  In any case its information
content can be investigated.  Either the dynamics of the remnant allows
the information to subsequently escape (in the extreme case the remnant
could immediately explode), or one could
directly probe the core to investigate its internal state.

The dependence of the core size (and hence mass) on the information
content also offers an escape from the objections to Planck-sized
remnants.  Such a core is similar to other types of macroscopic bodies
whose information content increases with size and mass.  For example,
since there are only finitely many states below a given
energy, the microcanonical ensemble is defined for large remnants, in
contrast to those of Planck size.
Therefore not only has one escaped the difficulty
with storing arbitrarily large amounts of information in a Planck-scale
remnant, but the problems of loops and thermodynamics are ameliorated as
well.
%

One possible causal structure for the process of black hole formation in
such a picture is shown in \nfig\two{Shown is a spacetime diagram in which a
collapsing body forms a black hole with internal core.  The black hole then
evaporates until the horizon and core meet.  The core then becomes a
massive remnant.  Some of the light cones are pictured also.  Note that this
diagram would be equally applicable in the case where the remnant is Planck
size; in that case the core and horizon would meet at scales comparable to
$\lpl$.}\nfig\three{Pictured is a Penrose diagram corresponding to \two.
Note that this diagram would also be equally applicable to the case of a
Planck-sized remnant.}\figs{\two,\three}.  The surface of the
remnant (in so far as it is sharply defined) travels on a spacelike, hence
acausal, trajectory inside the horizon.  However, the trajectory needn't be
spacelike outside the horizon.  Furthermore, this behavior is only
encountered when entropy densities become planckian.  One might also have
taken solace in the fact that with the curvature estimate for the core
radius,
the surface is encountered
only when curvatures become planckian; it is perhaps not surprising
to encounter acausal phenomena in such regimes.   However, if it
is assumed
that the core doesn't shrink with the mass, in accord with the idea
that it must stay large to accommodate the information, then one faces the
prospect of encountering the core before reaching planckian curvatures.
Therefore, even though it is cloaked by
the horizon, the spacelike propagation of the core surface is potentially
the most serious drawback of this proposal.  One might conjecture that
naked causality violation is censored in an attempt to live with this
proposal.
Alternatively, a possible
escape from this objection is mixing of black hole and white hole states.
The resulting causal structure is not easily pictured, and in any case this
possibility is equally speculative.\foot{Such mixing has however been
considered in a toy model of collapsing shells of matter\refs{\Haij}.}

It is an an empirical fact that each attempt to  resolve the
black hole information problem engenders apparently
objectionable behavior.
Whether this scenario is as
offensive as other proposed resolutions of the
information problem
is unclear.  The bottom line on the information problem is that information
is either lost, escapes during the Hawking process, remains after the
black hole reaches $M\sim\Mpl$, or emerges much earlier as a result of new
physics.
This paper is
advocating consideration of the latter possibility.

If such remnants in fact exist, then it should in principle be
possible to find them through direct observation.
One looks for massive high density objects\foot{For example, $M_{\rm
remnant} \sim \Mpl (M/\Mpl)^{2/3}$ and $\rho_{\rm remnant}\sim
(\Mpl/\lpl^3) (\Mpl/M)^{4/3}$ from the na\"\i ve entropy bound \ebound.} that
nonetheless do not have horizons and therefore may emit or scatter
radiation from their surfaces.  Furthermore, such objects could
clearly have astrophysical consequences, although determining these
consequences would depend both on the initial mass distribution of
black holes from which they formed as well as on their dynamics (\eg\
stability).
The latter in particular is uncertain due to the lack of
detailed knowledge about the short distance physics responsible for
their existence.
Possible implications for, \eg, dark matter remain to be
investigated.

In summary, with the assumption that there is an upper bound on
the amount of information that can be contained in a given volume, or
with other assumptions about a Planck scale cutoff in quantum
gravity, one is drawn to the conclusion that black holes have finite
sized cores and that these cores could become large, massive
remnants after Hawking
emission.  Such remnants offer an easy escape from the black hole
information problem implied by the traditional Hawking evaporation
scenario.  They also avoid objections raised to other attempts to
solve this problem.  However, they may encounter problems with causality.
They in any case must remain purely conjectural
until we have a sufficient understanding of short
distance gravity to definitively
predict their existence, or until they are observed.

\vglue1cm
\centerline{\bf ACKNOWLEDGMENTS}
I wish to thank Y. Aharonov, M. Bowick, S. Hawking, K. Kucha\v r,  and J.
Preskill
for conversations, G. Horowitz for comments on a draft,
and A.~Strominger for conversations and valuable comments.
L. Susskind and L. Thorlacious have considered
interpretation of static solutions\refs{\HawkETD\SuTh-\QBH}
of the model of \refs{\CGHS} in terms of
massive remnants; I thank them for conversations.
This work was supported in part by DOE OJI grant DE-FG03-91ER40168 and NSF
PYI grant PHY-9157463.

\listrefs
\listfigs
\end